\documentclass[numbers,sort,merge,square, comma, sort&compress, final,5p,times,twocolumn]{elsarticle}
\pdfoutput=1
\usepackage{atlasphysics}
\usepackage{float}
\usepackage{mathrsfs}
\usepackage{graphicx}
\usepackage{hyperref}
\usepackage{multirow}
\usepackage{rotating}
\usepackage{url}
\usepackage{textpos}
\usepackage{color}
\usepackage{epstopdf}
\usepackage{xspace}
\usepackage{lineno}
\biboptions{sort,merge}
\usepackage{preprintcover}  

\PreprintCoverPaperTitle{Search for displaced muonic lepton jets from light Higgs boson decay in
proton-proton collisions at \myrts~ = 7 ~\tev~ with the ATLAS detector}  
\PreprintIdNumber{CERN-PH-EP-2012-241}  
\PreprintCoverAbstract{A search is performed for collimated muon pairs displaced from the primary vertex produced in the decay of long-lived neutral particles in proton-proton collisions at $\sqrt{s}$ = 7 \tev~ centre-of-mass energy, with the ATLAS detector at the LHC. In a \intlumi event sample collected during 2011, the observed data are consistent with the Standard Model background expectations. Limits on the product of the production cross section and the branching ratio of a Higgs boson decaying to hidden-sector neutral long-lived particles are derived as a function of the particles' mean lifetime. }  
\PreprintJournalName{Pysics Letters B}  

\newcommand{\leptonjet}{Search for displaced muonic lepton jets from light Higgs boson decay in
proton-proton collisions at \myrts~ = 7 ~\tev~ with the ATLAS detector}
\newcommand{\maybesection}[1]{\par  \vspace{1.6mm} {{\bf #1}} \vspace{0.2mm}\par}

\newcommand{\pythia}{{\tt PYTHIA}}
\newcommand{\mcatnlo}{{\tt MC@NLO}}
\newcommand{\geant}{{\tt GEANT4}}
\newcommand{\bridge}{{\tt BRIDGE}}
\newcommand{\madgraph}{{\tt MadGraph}}

\newcommand{\mydelphi}{$\Delta \phi$}
\newcommand{\mydelr}{$\Delta R$\xspace}
\newcommand{\absetaleq}{$\mid\eta\mid$~$\leq$\xspace}
\newcommand{\intlumi}{1.9~fb$^{-1}$\xspace}

\newcommand{\higgsl}{$m_{H}=~$100 GeV\xspace}
\newcommand{\higgsh}{$m_{H}=~$140 GeV\xspace}
\newcommand{\higgshnos}{$m_{H}=~$140 GeV}
\newcommand{\gammad}{$\gamma_{d}$\xspace}
\newcommand{\gdmumu}{\gammad$\rightarrow \mu~\mu$\xspace}
\newcommand{\higgsd}{{\it H}$\rightarrow \gamma_{d}~\gamma_{d}+ X$\xspace}
\newcommand{\mytev}{TeV\xspace}

\newcommand{\myrts}{$\sqrt{s}$\xspace}
\newcommand{\isol}{$E_{\mathrm{T}}^{\mathrm{isol}}$\xspace}
\newcommand{\ctau}{c$ \tau$\xspace}
\newcommand{\sigmabr}{$\sigma\times$BR }
\newcommand{\SMs}{\small SM}
\newcommand{\stats}{\small stat}
\newcommand{\twoROIf}{\footnotesize 2ROI}
\newcommand{\twoROIt}{\tiny 2ROI}
\newcommand{\oneROIf}{\footnotesize 1ROI}
\newcommand{\TPf}{\footnotesize TP}
\newcommand{\TPss}{\scriptsize TP}
\newcommand{\MJf}{\footnotesize MJ}
\newcommand{\recf}{\footnotesize rec}
\newcommand{\recss}{\scriptsize rec}
\newcommand{\MJss}{\scriptsize MJ}
\newcommand{\pTsquare}{$p_{\mathrm{T}}^{\mathrm{2}}$\xspace}
\newcommand{\SumpT}{$ \Sigma p_{\mathrm{T}}^{\mathrm{ID}}$\xspace}
\def\Zmumu{\ensuremath{Z \rightarrow \mu\mu}}
\makeatletter
\g@addto@macro\bfseries{\boldmath}
\makeatother
\begin{document}
\begin{frontmatter}
\title{\leptonjet}
\author{The ATLAS Collaboration}

\begin{abstract}
A search is performed for collimated muon pairs displaced from the primary vertex produced in the decay of long-lived neutral particles in proton-proton collisions at $\sqrt{s}$ = 7 \tev~ centre-of-mass energy, with the ATLAS detector at the LHC. In a \intlumi event sample collected during 2011, the observed data are consistent with the Standard Model background expectations. Limits on the product of the production cross section and the branching ratio of a Higgs boson decaying to hidden-sector neutral long-lived particles are derived as a function of the particles' mean lifetime. 
\end{abstract}
\end{frontmatter}
\begin{figure}[ht!]
\centering
\includegraphics[width=55mm]{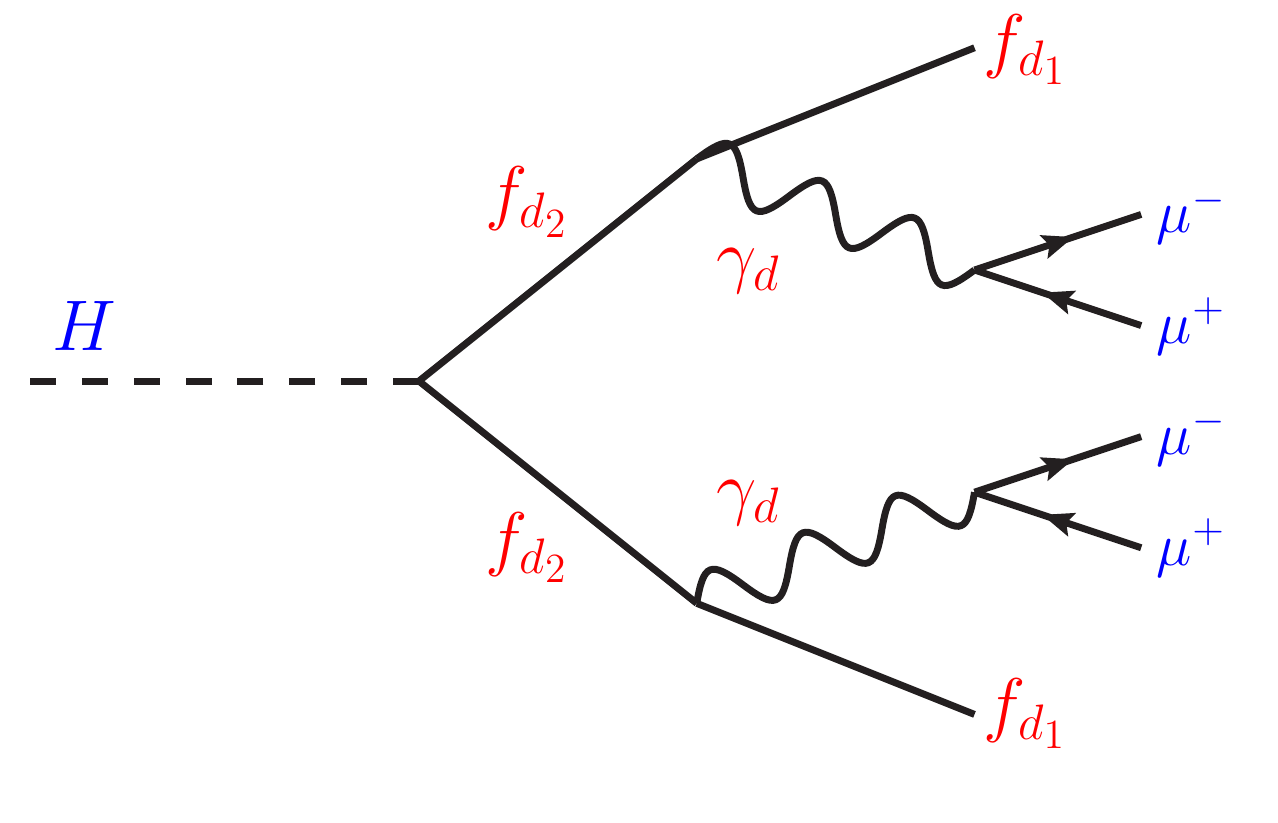}
\caption{Schematic picture of the Higgs boson decay chain, H$\rightarrow$2($f_{d2}\rightarrow f_{d1}$\gammad). The Higgs boson decays to two hidden fermions ($f_{d2}$). Each hidden fermion decays to a \gammad and to a stable hidden fermion ($f_{d1}$), resulting in two muon jets from the \gammad decays in the final state.}
\label{fig:model}
\end{figure}

%
\maybesection{1. Introduction}
\vspace{0.1cm}
A search is presented for long-lived neutral particles decaying to final states containing collimated muon pairs in proton-proton collisions at $\sqrt{s}$ = 7 \tev~ centre-of-mass energy. The event sample, collected during 2011 at the LHC with the ATLAS detector, corresponds to an integrated luminosity of  \intlumi. The model considered in this analysis consists of a Higgs boson decaying to a new hidden sector of particles which finally produce two sets of collimated muon pairs, but the search described is equally valid for other, distinct models such as heavier Higgs boson doublets, singlet scalars or a $Z^\prime$ that decay to a hidden sector and eventually produce collimated muon pairs. \\
Recently, evidence for the production of a boson with a mass of about 126~\gev~has been published by ATLAS~\cite{higgsatl} and CMS~\cite{higgscms}. The observation is compatible with the expected production and decay of the Standard Model (SM) Higgs boson~\cite{HIGGS1,HIGGS2,HIGGS3} at this mass. Testing the SM Higgs hypothesis is currently of utmost importance. To this end two effects may be considered:
 (i) additional resonances which arise in an extended Higgs sector found in many extensions of the SM, or
 (ii) rare Higgs boson decays which may deviate from those predicted by the SM. In this Letter we search for a scalar that decays to a light hidden sector, focusing on the 100 GeV to 140 GeV mass range. In doing so, we cover both of the above aspects, deriving constraints on additional Higgs-like bosons, as well as placing bounds on the branching ratio of the discovered 126 GeV resonance into a hidden sector of the kind described below. \\
The phenomenology of light hidden sectors has been studied extensively over the past few years~\cite{b1,b2,b3,b4,b5}. Possible characteristic topological signatures of such extensions of the SM are ``lepton jets". A lepton jet is a cluster of highly collimated particles: electrons, muons and possibly pions~\cite{b2,b6,b7,b8}. These arise if light unstable particles with masses in the \mev~to~\gev~range (for example dark photons, \gammad ) reside in the hidden sector and decay predominantly to SM particles. At the LHC, hidden-sector particles may be produced with large boosts, causing the visible decay products to form jet-like structures. Hidden-sector particles such as \gammad may be long-lived, resulting in decay lengths comparable to, or larger than, the detector dimensions. The production of lepton jets can occur through various channels. For instance, in supersymmetric models, the lightest visible superpartner may decay into the hidden sector.  Alternatively, a scalar particle that couples to the visible sector may also couple to the hidden sector through Yukawa couplings or the scalar potential. This analysis is focused on the case where the Higgs boson decays to the hidden sector~\cite{b9, b10}. The SM Higgs boson has a narrow width into SM final states if $m_{H} < 2 m_W$. Consequently, any new (non-SM) coupling to additional states, which reside in a hidden sector, may contribute significantly to the Higgs boson decay branching ratios. Even with new couplings, the total Higgs boson width is typically small, well below the order of one GeV. If a SM-like Higgs boson is confirmed, it will remain important to constrain possible rare decays, e.g. into lepton jets. \\
Neutral particles with large decay lengths and collimated final states represent, from an experimental point of view, a challenge both for the trigger and for the reconstruction capabilities of the detector. Collimated particles in the final state can be hard to disentangle due to the finite granularity of the detectors; moreover, in the absence of inner tracking detector information and a primary vertex constraint, it is difficult to reconstruct charged-particle tracks from decay vertices far from the interaction point (IP). 
The ATLAS detector~\cite{ATLASTDR} is equipped with a muon spectrometer (MS) with high-granularity tracking detectors that allow charged-particle tracks to be reconstructed in a standalone configuration using only the muon detector information (MS-only). This is a crucial feature for detecting muons not originating from the primary interaction vertex. \\
The search presented in this Letter focuses on neutral particles decaying to the simplest type of muon jets (MJs), containing only two muons; prompt MJ searches have been performed both at the Tevatron~\cite{tevatron1,tevatron2} and at the LHC~\cite{CMS}. Other searches for displaced decays of a light Higgs boson to heavy fermion pairs have also been performed at the LHC~\cite{Hiddenv}. \\
The benchmark model used for this analysis is a simplified scenario where the Higgs boson decays to a pair of neutral hidden fermions ($f_{d2}$) each of which decays to one long-lived \gammad and one stable neutral hidden fermion ($f_{d1}$) that escapes the detector unnoticed, resulting in two lepton jets from the \gammad decays in the final state (see Fig.~\ref{fig:model}). The mass of the \gammad (0.4~\gev) is chosen to provide a sizeable branching ratio to muons~\cite{b9}.

%
\vspace{0.3cm}
\maybesection{2. The ATLAS Detector}
\vspace{0.1cm}
ATLAS is a multi-purpose detector~\cite{ATLASTDR} at the LHC, consisting of an inner tracking system (ID) embedded in a superconducting solenoid, which provides a 2~T magnetic field parallel to the beam direction, electromagnetic and hadronic calorimeters and a muon spectrometer using three air-core toroidal magnet systems\footnote{ ATLAS uses a right-handed coordinate system with its origin at the nominal interaction point (IP) in the centre of the detector and the $z$-axis coinciding with the beam pipe axis. The $x$-axis points from the IP to the centre of the LHC ring, and the $y$-axis points upward. Cylindrical coordinates ($r$,$\phi$) are used in the transverse plane, $\phi$ being the azimuthal angle around the beam pipe. The pseudorapidity is defined in terms of the polar angle $\theta$ as \mbox{$\eta$ = $-$ln tan($\theta$/2)}.}. 
The trigger system has three levels~\cite{L1TRIG} called Level-1 (L1), Level-2 (L2) and Event Filter (EF).  L1 is a hardware-based system using information from the calorimeter and muon spectrometer, and defines one or more Regions of Interest
(ROIs), geometrical regions of the detector, identified by ($\eta$, $\phi$) coordinates, containing interesting physics objects.
L2 and the EF (globally called the High Level Trigger, HLT) are software-based systems and can access information from all sub-detectors. 
The ID, consisting of silicon pixel and micro-strip detectors and a straw-tube tracker, provides precision tracking of charged particles for \mbox{\absetaleq~2.5}. The electromagnetic and hadronic calorimeter system covers \mbox{\absetaleq~4.9} and, at \mbox{$\eta$~=~0}, has a total depth of 9.7 interaction lengths (22 radiation lengths in the electromagnetic part). The MS provides trigger information (\mbox{\absetaleq~2.4}) and momentum measurements (\mbox{\absetaleq~2.7}) for charged particles entering the spectrometer. It consists of one barrel and two endcap parts, each with 16 sectors in $\phi$, equipped with precision tracking chambers and fast detectors for triggering. Monitored drift tubes are used for precision tracking in the region \mbox{\absetaleq~2.0} and cathode strip chambers are used for 2.0 $\leq$ \mbox{\absetaleq~2.7}. The MS detectors are arranged in three stations of increasing distance from the IP: inner, middle and outer. The air core toroidal magnetic field allows an accurate charged particle reconstruction independent of the ID information. The three planes of trigger chambers (resistive plate chambers in the barrel and the thin gap chambers in the endcaps) are located in middle and outer (only in the barrel) stations. The L1 muon trigger requires hits in the middle stations to create a low tranverse momentum (\pt) muon ROI or hits in both the middle and outer stations for a high \pt~ROI. 
The muon ROIs have a spatial extent of \mbox{0.2$\times$0.2} (\mbox{$\Delta\eta\times\Delta\phi$}) in the barrel and of \mbox{0.1$\times$0.1} in the endcap. 
L1 ROI information seeds, at HLT level, the reconstruction of muon momenta using the precision chamber information. In this way sharp trigger thresholds up to 40 \gev~can be obtained.
%
\vspace{0.1cm}
\maybesection{3. Signal and background simulation}
\vspace{0.1cm}
The set of parameters used to generate the signal Monte Carlo samples is listed in Table~\ref{tab:param}. The Higgs boson is generated through the gluon-gluon fusion production mechanism which is the dominant process for a low mass Higgs boson. The gluon-gluon fusion Higgs boson production cross section in {\it pp} collisions at \myrts = 7 \tev, estimated at the next-to-next-to-leading order (NNLO)~\cite{HiggsCrossS}, is $\sigma_{\textrm\SMs} = $ 24.0 pb for $m_{H}=$~100~\gev~and $\sigma_{\textrm\SMs} = $ 12.1 pb for $m_{H}=$~140~\gev. The \pythia~generator~\cite{PYTHIA} is used, linked together with \madgraph 4.4.2~\cite{b12} and \bridge~\cite{BRIDGE}, for gluon-gluon fusion production of the Higgs boson and the subsequent decay to hidden-sector particles. \\
As discussed in the introduction, the signal is chosen to enable a study of rare, non-SM, Higgs boson decays in the (possibly extended) Higgs sector.   To do so we choose two points which envelope a mass range covering the 126 GeV resonance. The lower mass point, $m_H=$ 100 GeV, is chosen to be compatible with the decay-mode-independent search by OPAL at LEP~\cite{opal}. The higher mass point, $m_H=$ 140 GeV, is chosen well below the $WW$ threshold, where a sizeable branching ratio into a hidden sector may be naturally achieved. The masses of $f_{d2}$ and $f_{d1}$ are chosen to be light relative to the Higgs boson mass, and far from the kinematic threshold at $m_{f_{d1}} + m_{\gamma_{d}} = m_{f_{d2}}$. For the chosen dark photon mass (0.4~\gev), the \gammad decay branching ratios are expected to be~\cite{b9}: 45$\%~ \ee$, 45$\%~\mu^+\mu^-$, 10$\%~\pi^+\pi^-$. Thus $20\%$ of the Higgs \higgsd decays are expected to have the required four-muon final state. \\
The mean lifetime $\tau$ of the \gammad (expressed throughout this Letter as $\tau$ times the speed of light $c$) is a free parameter of the model. In the generated samples \ctau is chosen so that a large fraction of the decays occur inside the sensitive ATLAS detector volume, i.e.\ up to 7 m in radius and 13 m along the $z$-axis, where the trigger chambers of the middle stations are located. The detection efficiency can then be estimated for a range of \gammad mean lifetime through re-weighting of the generated samples. \\

\begin{table}[H]
\small
\centering
\begin{tabular}{|c c c c c|}\hline
Higgs mass & $m_{f_{d2}}$ & $m_{f_{d1}}$  & \gammad mass  & \ctau  \\
$[\gev]$  &$[\gev]$ &$[\gev]$ & $[\gev]$ & [mm] \\\hline
100& 5.0 & 2.0 & 0.4 & 47\\\hline
140 & 5.0 & 2.0 & 0.4 & 36 \\\hline
\end{tabular}
\caption{Parameters used for the Monte Carlo simulation. The last column is the \gammad mean lifetime $\tau$ multiplied by the speed of light $c$, expressed in mm.}
\label{tab:param}
\end{table} 

\begin{figure}[ht!]
\centering
\includegraphics[width=70mm]{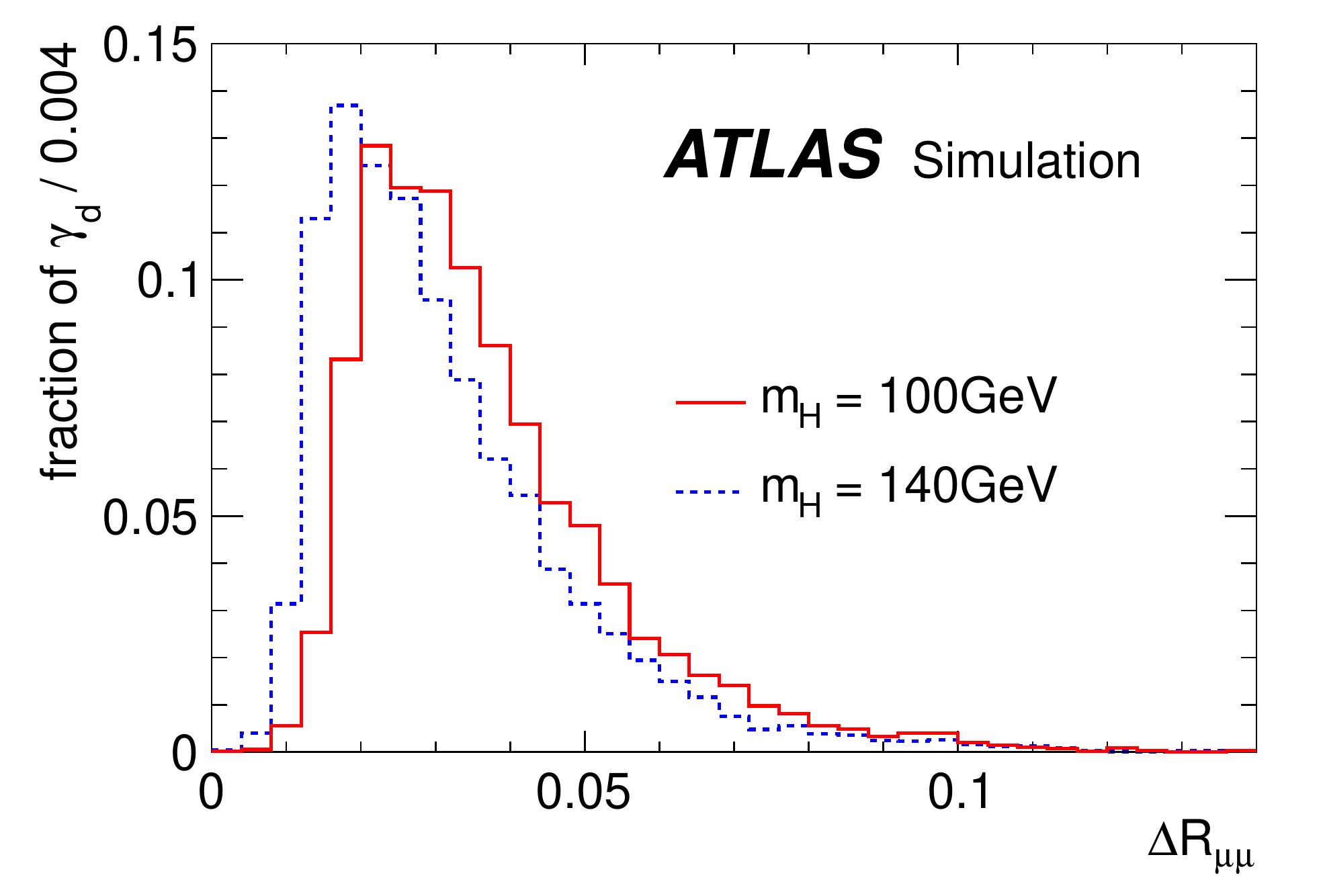}
\caption{$\Delta R$ distribution between the two muons from the $\gamma_{d}$ decay for the signal Monte Carlo samples with $m_{H}$ = 100 \gev~and $m_{H}$ = 140 \gev.}
\label{fig:drmu}
\end{figure}

Potential backgrounds include all the processes which lead to real prompt muons in the final state such as the SM processes {\it W}+jets, {\it Z}+jets, \ttbar, {\it WW}, {\it WZ}, and {\it ZZ}. However, the main contribution to the background is expected from processes giving a high production rate of secondary muons which do not point to the primary vertex, such as decays in flight of $K/\pi$ and heavy flavour decays in multi-jet processes, or muons due to cosmic rays. The prompt lepton background samples are generated using \pythia~({\it W}+jets, and {\it Z}+jets) and \mcatnlo~\cite{mcatlno}~(\ttbar, {\it WW}, {\it WZ}, and {\it ZZ}).
 The generated Monte Carlo events are processed through the full ATLAS simulation chain based on \geant~\cite{GEANT4, ATLSIM}. Additional {\it pp} interactions in the same and nearby bunch crossings (pile-up) are included in the simulation. All Monte Carlo samples are re-weighted to reproduce the observed distribution of the number of interactions per bunch crossing in the data.  For the multi-jet background evaluation a data-driven method is used. The cosmic-ray background is also evaluated from data.
%
\vspace{0.4cm}
\maybesection{4. The kinematics of the signal}
\vspace{0.1cm}
The main kinematic characteristics of the signal sample are:
\begin{itemize}
\item{The \gammad pair are emitted approximately back-to-back in $\phi$, with an angular spread of the distribution due to the emission of the $f_{d1}$. }
\item{The average \pt~of the $\gamma_{d}$ in the laboratory frame is about 20 \gev~for \higgsl and 30 \gev~for \higgsh; due to the small mass of the \gammad, large boost factors in the decay length should be expected.}
\item{ Fig.~\ref{fig:drmu} shows the distribution of $\Delta R = \sqrt{(\Delta\eta)^2 + (\Delta\phi)^2}$ between the two muons  from the \gammad decay. The $\Delta R$ is computed at the decay vertex of the \gammad from the vector momenta of the two muons. Due to the small mass of the \gammad the $\Delta R$ is almost always below 0.1.}
\end{itemize}
Since the two $f_{d1}$ are, like the two \gammad, emitted back-to-back in $\phi$, the observed missing transverse momentum \MET, computed at the event-generator level, is small and cannot be used as a discriminating variable against the background.

\begin{figure}[ht!]
\centering
\includegraphics[width=70mm]{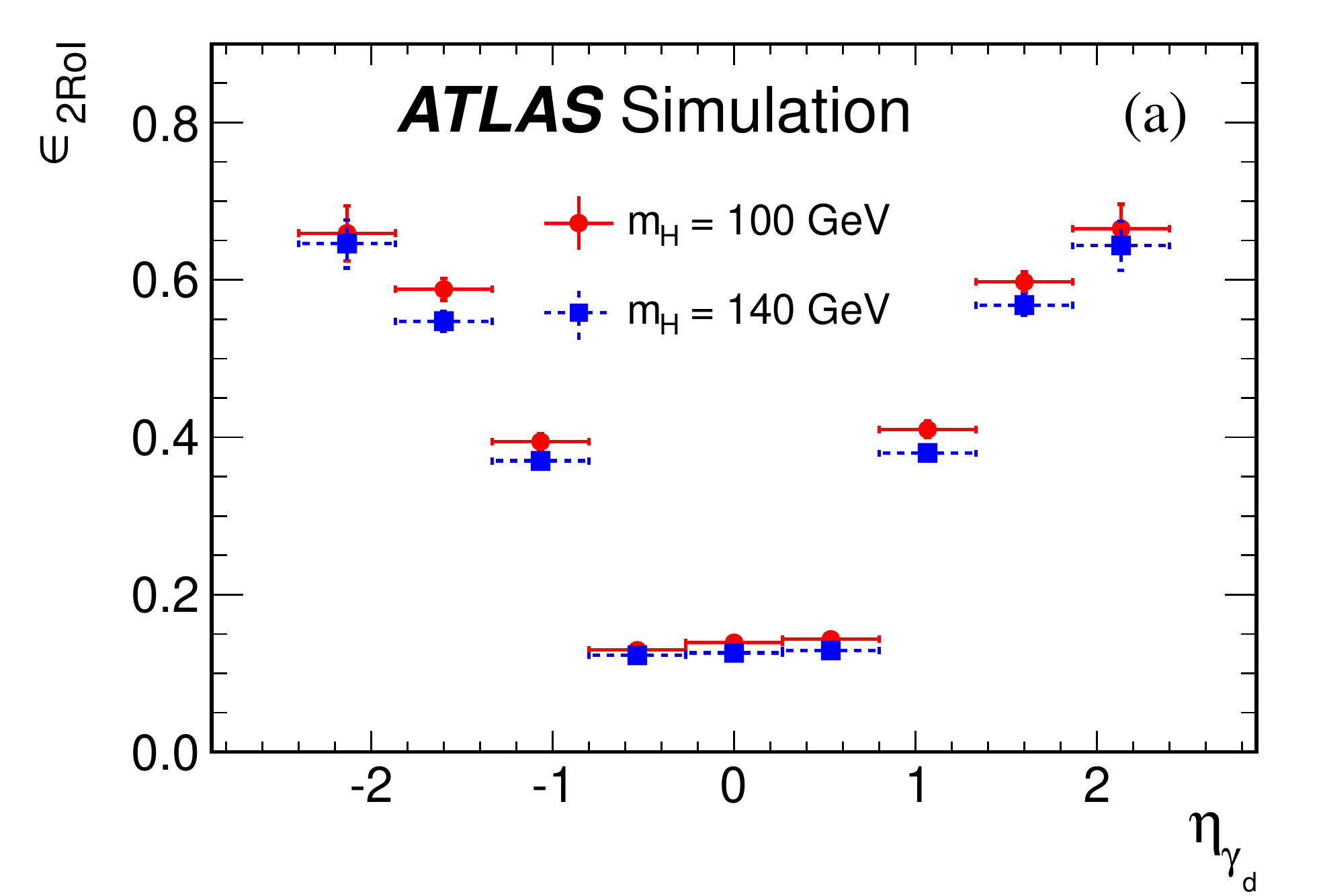}
\includegraphics[width=70mm]{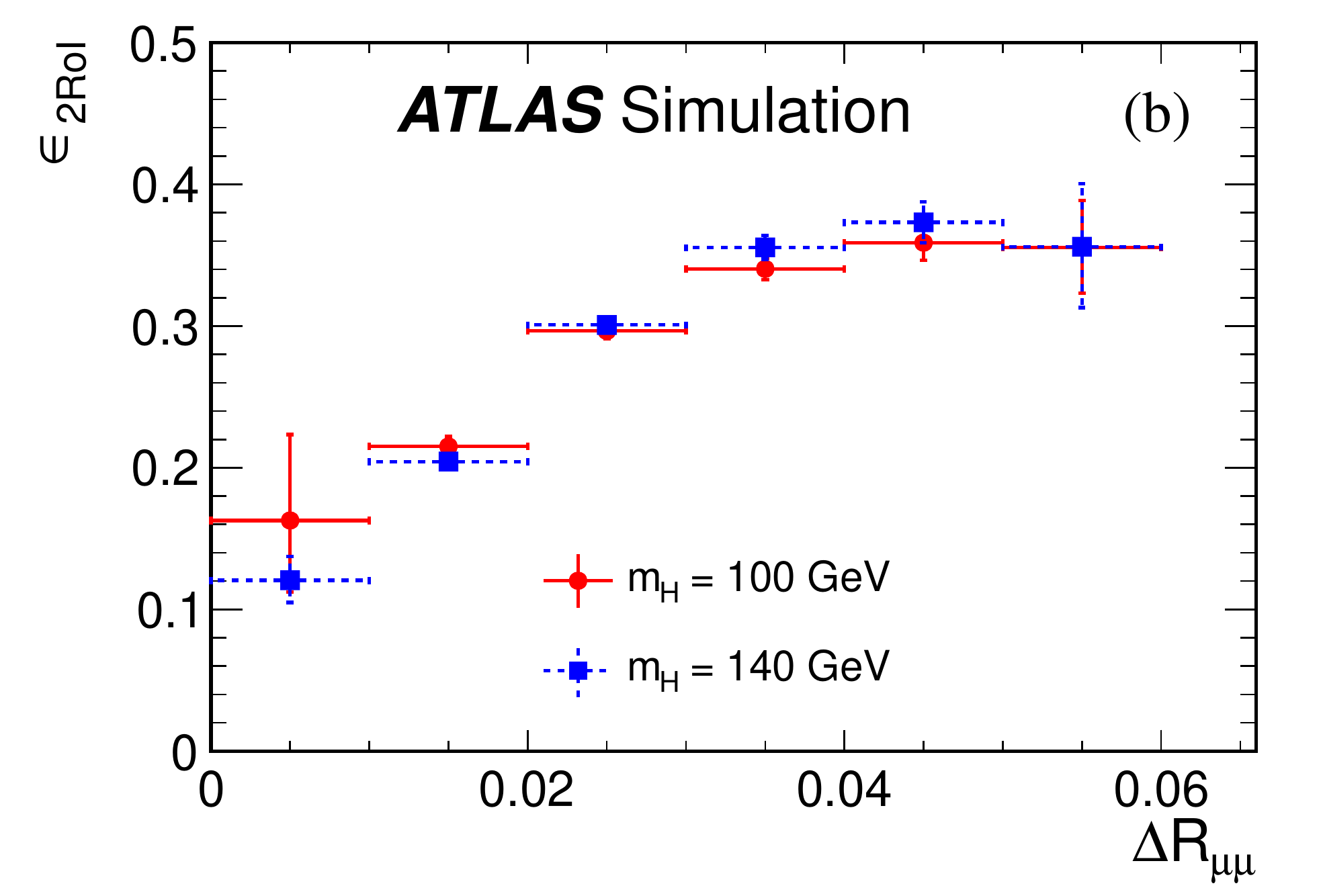}
\caption{$\varepsilon_{\textrm\twoROIt}$ as a function (a) of the $\eta$ of the \gammad and (b) of the \mydelr of the muon pair for the Monte Carlo samples with Higgs boson masses of 100 \gev~and 140 \gev. The errors are statistical only.}
\label{fig:treggeffVSetadr2}
\end{figure}

\vspace{0.4cm}
\maybesection{5. Data samples and trigger selection}
The dataset used for this analysis was collected at a centre-of-mass energy of 7 \tev~during the first part of 2011, where a low level of pile-up events in the same bunch-crossing was present (an average of $\approx 6$ interactions per crossing). Only periods in which all ATLAS subdetectors were operational are used. The total integrated luminosity used is 1.94~$\pm$~$0.07$~fb$^{-1}$~\cite {LUMI1,LUMI2}. All events are required to have at least one reconstructed vertex along the beam line with at least three associated tracks, each with \pt~$\geq$ 0.4 \GeV. The primary interaction vertex is defined to be the vertex whose constituent tracks have the largest $\Sigma$\pTsquare. This analysis deals with displaced \gammad decays with final states containing only muons. Signal events are therefore characterized by a four-muon final state with the four muons coming from two displaced decay vertices. Due to the relatively low \pt~of the muons and to the displaced decay vertex, a low-\pt~multi-muon trigger with muons reconstructed only in the MS is needed. In order to have an acceptably low trigger rate at a low  \pt~threshold, a multiplicity of at least three muons is required. Candidate events are collected using an unprescaled HLT trigger with three reconstructed muons of \pt~$ \geq $ 6 \GeV, seeded by a L1-accept with three different muon ROIs. These muons are reconstructed only in the MS, since muons originating from a neutral particle decaying outside the pixel detector will not have a matching track in the ID tracking system. The trigger efficiency for the Monte Carlo signal samples, defined as the fraction of events passing the trigger requirement with respect to the events satisfying the analysis selection criteria (described in Section 6) is 0.32$\pm 0.01_{\textrm\stats}$ for \higgsl and 0.31$\pm 0.01_{\textrm\stats}$ for \higgsh. \\
The main reason for the relatively low trigger efficiency is the small opening $\Delta R$ between the two muons of the \gammad decay ($\Delta R  \leq 0.1$) shown in Fig.~\ref{fig:drmu}. These values of  $\Delta R$ are often smaller than the L1 trigger granularity; in this case the L1 produces only one ROI. The trigger only fires if at least one of the \gammad produces two distinct L1 ROIs. The single \gammad ROI efficiency, $\varepsilon_{\textrm\twoROIf}$ ($\varepsilon_{\textrm\oneROIf}$), defined as the fraction of \gammad passing the offline selection that give two (one) trigger ROIs is 0.296~$\pm~0.004_{\textrm\stats}$ (0.626~$\pm~0.004_{\textrm\stats}$) for \higgsl and 0.269~$\pm~0.003_{\textrm\stats}$ (0.653~$\pm~0.003_{\textrm\stats}$) for \higgsh. Fig.~\ref{fig:treggeffVSetadr2} shows the $\varepsilon_{\textrm\twoROIf}$ as a function of the dark photon $\eta$ and of the \mydelr of the two muons from the \gammad decay. 
The increased trigger granularity in the endcap and the efficiency decrease  at small values of \mydelr are clearly visible. \\
The systematic uncertainty on the trigger efficiency is estimated with a sample of \Jmumu~from collision data and a corresponding sample of Monte Carlo events, using the tag-and-probe (TP) method. A cut on $\Delta R  \leq 0.1$ between the two muons is used to reproduce the small track-to-track spatial separation in the MS of the signal. The tag is a (MS+ID) combined muon, defined as a MS-reconstructed muon that is associated with a trigger object and combined with a matching ``good ID track". Good ID tracks must have at least one hit in the pixel detector, at least six hits in the silicon micro-strip detectors and at least six hits in the straw-tube tracker. The probe is a good ID track which, when combined with the tag track, gives an invariant mass inside a 100 \mev~window around the \jpsi~mass. A muon ROI that matches the probe in $ \eta$ and $\phi$, and is different from the ROI associated with the tag, is searched for. The number of probes with a matched ROI divided by the number of probes without a matched ROI gives the $\varepsilon$$_{\textrm\twoROIf}^{\textrm\TPf}$/$\varepsilon$$_{\textrm\oneROIf}^{\textrm\TPf}$ ratio. Values of $\varepsilon$$_{\textrm\twoROIf}^{\textrm\TPf}$/$\varepsilon$$_{\textrm\oneROIf}^{\textrm\TPf}$ = 0.42$\pm 0.05_{\textrm\stats}$ for the \Jmumu~data and $\varepsilon$$_{\textrm\twoROIf}^{\textrm\TPf}$/$\varepsilon$$_{\textrm\oneROIf}^{\textrm\TPf}$ = 0.39$\pm 0.05_{\textrm\stats}$ for the corresponding Monte Carlo sample are obtained. The relative statistical uncertainty on the difference between these two estimates is 17$\%$ and this is taken conservatively to be the systematic uncertainty on the trigger efficiency.

%
\vspace{0.4cm}
\maybesection{6. Muon Jets reconstruction and event selection}
\vspace{0.1cm}
 MJs from displaced \gammad decays are characterized by a pair of muons in a narrow cone, produced away from the primary vertex of the event. Consequently tracks reconstructed in the MS with a good quality track fit~\cite{reco} are used. MJs are identified using a simple clustering algorithm that associates all the muons in cones of $\Delta R = 0.2$, starting with the muon with highest \pt. The size of the cone takes into account the multiple scattering of the muons in the calorimeters. All the muons found in the cone are associated with a MJ. After this procedure, if any muons are unassociated with a MJ the search is repeated for this remainder, starting again with the highest \pt~muon. This continues until all possible MJs are formed. The MJ direction and momentum are obtained from the vector sum over all muons in the MJ. Only MJs with two reconstructed muons are accepted and only events with two MJs are kept for the subsequent analysis. \\
The possible contribution to the background of SM processes which lead to real prompt muon pairs in the final state is evaluated using simulated samples. After the trigger and the requirement of having two MJs in the event, their contributions have been found to be negligible. The only significant background sources are expected to be from processes giving a high production rate of secondary muons which do not point to the primary vertex, such as decays in flight of $K/\pi$ and heavy flavour decays in multi-jet production, or cosmic-ray muons not pointing to the primary vertex. \\
In order to separate the signal from the background, a number of discriminating variables have been studied. The multi-jet background can be significantly reduced by using calorimeter isolation requirements around the MJ direction. The calorimetric isolation variable \isol is defined as the difference between the transverse calorimetric energy \et~in a cone of $\Delta R = 0.4$ around the highest \pt~muon of the MJ and the \et~in a cone of $\Delta R = 0.2$; a cut \isol $ \leq$ 5$~ \gev~$ keeps almost all the signal. The isolation modelling is validated for real isolated muons with a sample of muons coming from $Z \rightarrow \mu\mu$ decays. To further improve the signal-to-background ratio, two additional discriminating variables are used: \mydelphi~ between the two MJs and \SumpT for the MJ, defined as the scalar sum of the transverse momentum of the tracks, measured in the ID, inside a cone $\Delta R = 0.4$ around the direction of the MJ. The muon tracks of the MJ in the ID, if any, are not removed from the isolation sum, so that prompt muons, which give a reconstructed track in both the ID and MS, will contribute to the \SumpT. As a consequence a cut on \SumpT of a few \gev~will remove prompt MJs or MJs with very short decay length. 

\begin{figure*}[ht!]
\centering
\includegraphics[width=60mm]{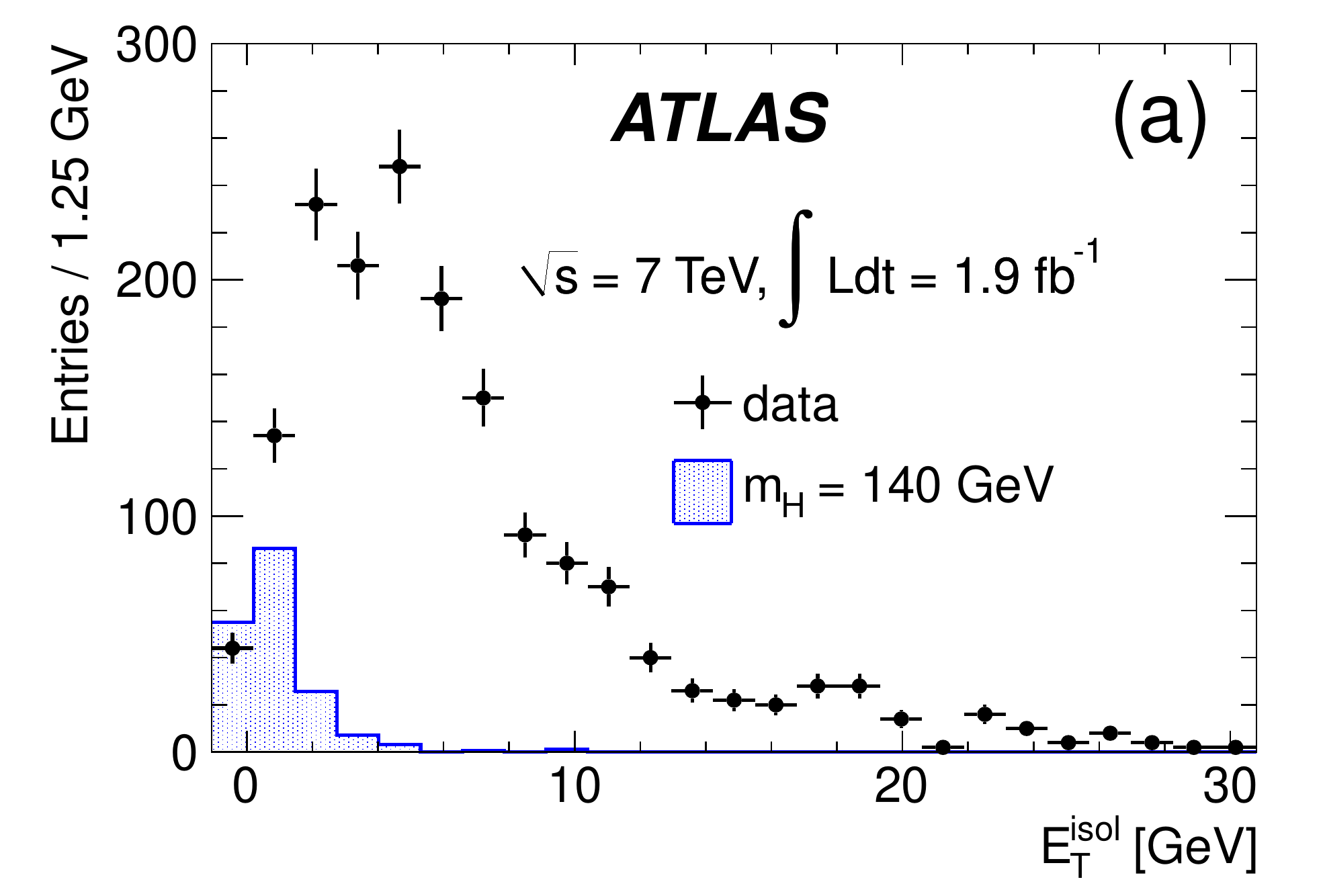}
\includegraphics[width=60mm]{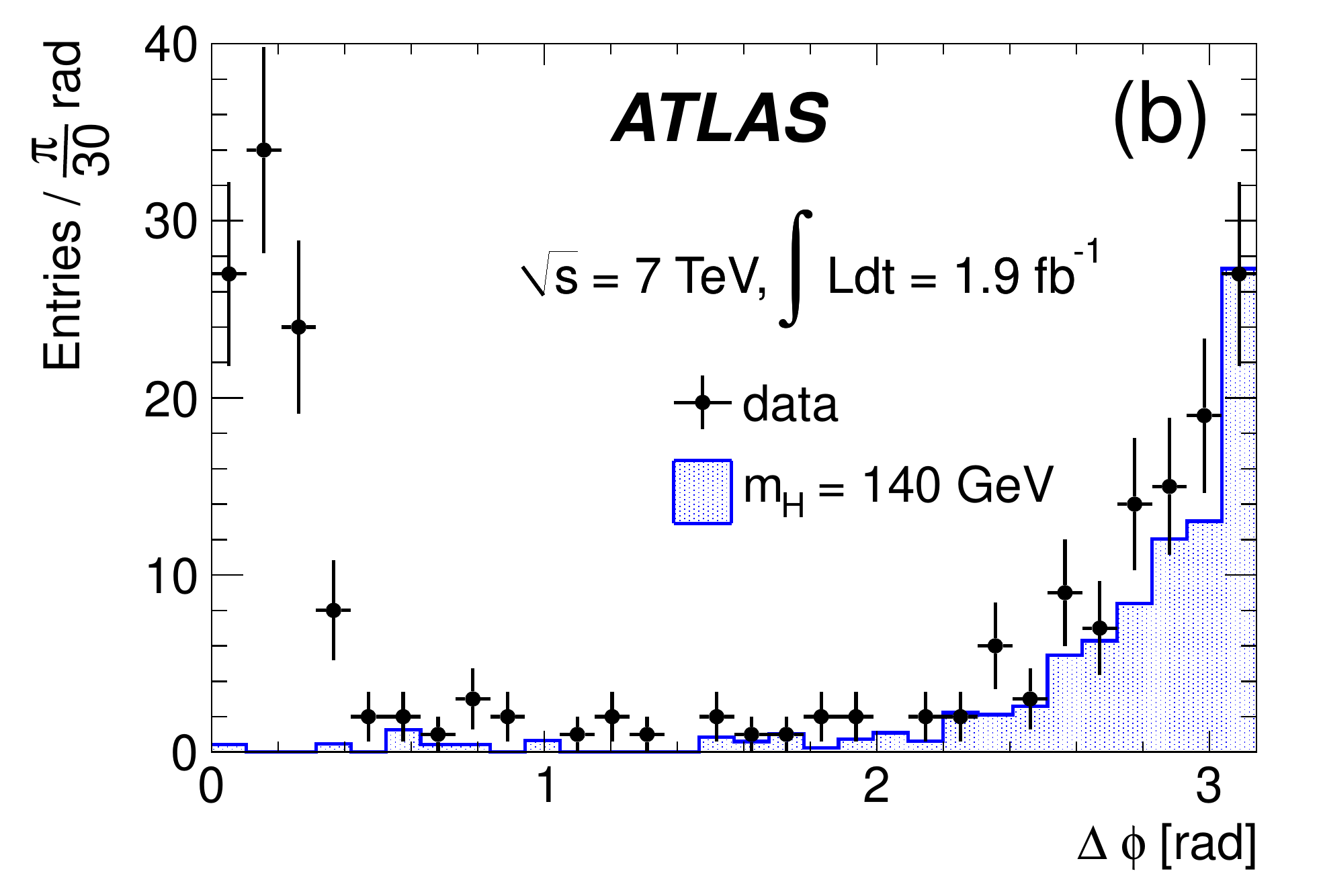}
\includegraphics[width=60mm]{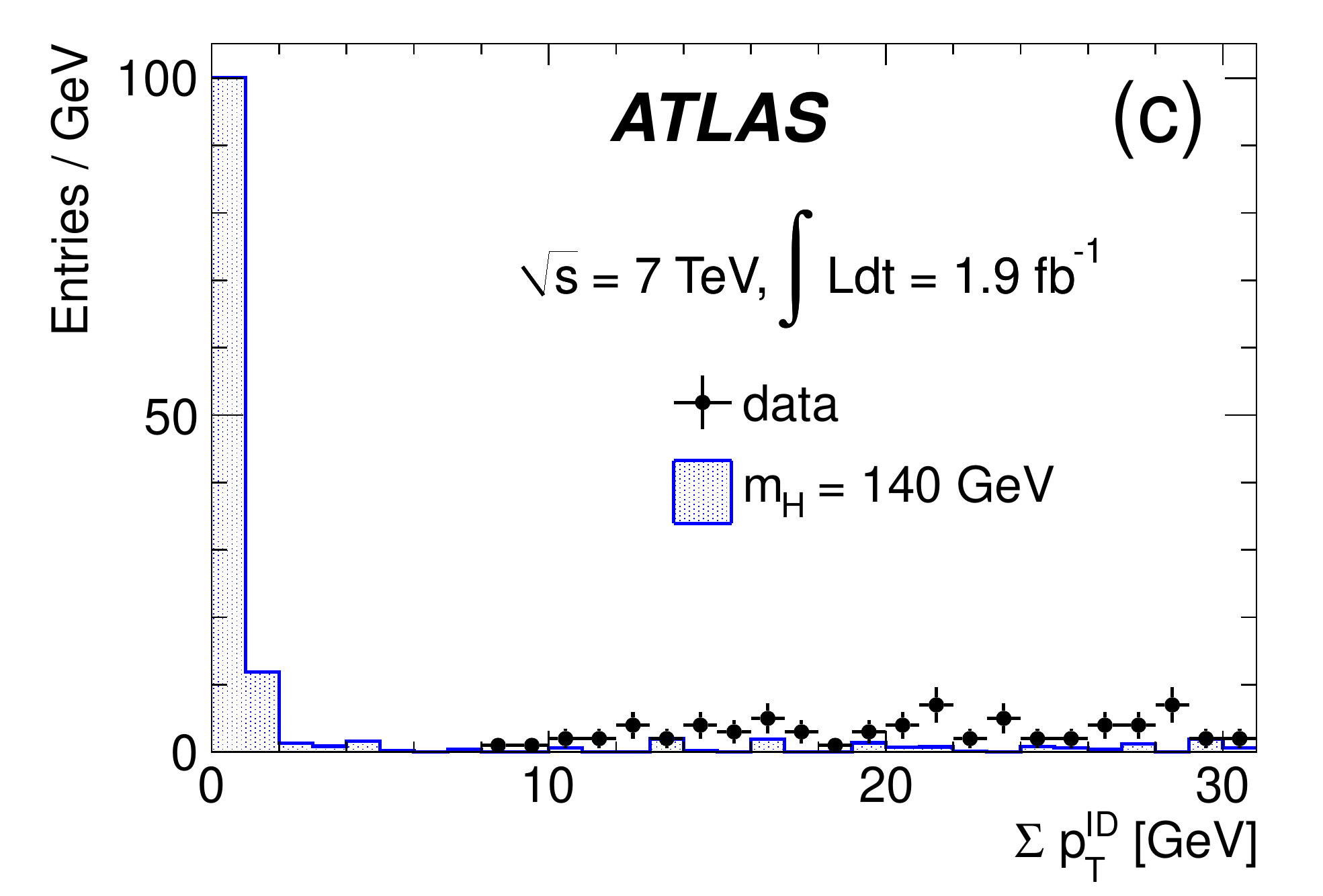}
\caption{Control plots for the cut variables on Monte Carlo (\higgsh) and on data. (a) Distribution of the calorimetric isolation around the MJ direction \isol after the requirement of two MJs in the event. (b) Distribution of $\Delta \phi$ between the two MJs after the requirement of  the isolation cut. (c) Distribution of \SumpT of the MJ after the requirement of the impact parameters cut. The points show the data and the histogram is the signal Monte Carlo normalized to \intlumi. The uncertainties are statistical only.}
\label{fig:sumpt_scan1}
\end{figure*}

For the background coming from cosmic-ray muons (mainly pairs of almost parallel cosmic-ray muons crossing the detector) a cut on the impact parameters of the muon tracks with respect to the primary interaction vertex is used. \\
The final set of selection criteria used is the following:
\begin{itemize}
\item Topology cut: events are required to have exactly two MJs, $N_{\textrm\MJf}$~= 2. 
\item MJ isolation: require MJ isolation with \isol $ \leq$ 5$~ \gev~$ for both MJs in the event.
\item Require $|\Delta \phi|$ $\geq$ 2 between the two MJs.
\item Require opposite charges for the two muons in a MJ (Q$_{\textrm\MJf} = 0$).
\item Require a cut on the transverse and longitudinal impact parameters of the muons with respect to the primary vertex: $|d_0| <$ 200 mm and $|z_0|< $ 270 mm.
\item Require \SumpT $< 3 \GeV$ for both MJs.
\end{itemize}
The distributions of the relevant variables at the different steps of the cut flow are shown in Fig.~\ref{fig:sumpt_scan1}. The results are summarized in Table~\ref{tab:cutflowABCD}. No events survive the selection in the data sample whereas the expected signals from Monte Carlo simulation, assuming 100$\%$ branching ratio for \higgsd and the parameters given in Table~\ref{tab:param}, are 75 or 48 events for Higgs boson masses of 100 \GeV~and 140 \GeV~respectively. The method used to estimate the cosmic-ray and multi-jet background yields, quoted in Table~\ref{tab:cutflowABCD}, is discussed in Section 7. 

\renewcommand{\arraystretch}{1.6}
\begin{table*}[ht!]
\centering
\footnotesize
\begin{tabular}{|c|c|c|c|c|c|c|}
\hline cut & cosmic-rays & multi-jet & total background &  \higgsl & \higgsh & data \\
\hline $N_{\textrm\MJss}= 2 $ &  $3.0\pm 2.1 $  & N/A  & N/A  & 135$\pm11_{-21}^{+29}$  & 90$\pm9_{-13}^{+17}$ & 871 \\
\hline  \isol $ \leq$ 5$~\gev$  &  $3.0\pm 2.1 $  &  N/A  & N/A  & 132$\pm11_{-21}^{+28}$ &  88$\pm9_{-13}^{+17}$ & 219 \\
\hline  $|\Delta \phi|\geq$ 2 &  $1.5\pm 1.5 $  & 153 $\pm$ 18 $\pm$ 9 &  155 $\pm$  18 $\pm$ 9 & 123$\pm11_{-19}^{+26}$ &  81$\pm9_{-12}^{+15}$& 104 \\
\hline  Q$_{\textrm\MJss}$ = 0 &  $1.5 \pm 1.5$  & 57 $\pm$15$\pm$22 & 59 $\pm$  15 $\pm$ 22  & 121$\pm11_{-19}^{+26}$  &  79$\pm8_{-12}^{+15}$ & 80 \\
\hline   $|d_0|$, $|z_0|$ &  $0_{-0} ^{+1.64}$ & 111$\pm$39$\pm$63 & 111$\pm$39$\pm$63 &  105$\pm10_{-16}^{+22}$ & 66$\pm8_{-10}^{+12}$ &  70 \\
\hline   \SumpT $< 3~\GeV$ &  $0_{-0} ^{+1.64}$  &  0.06$\pm 0.02 ^{+0.66} _{-0.06}$ & $0.06^{+1.64 +0.66} _{-0.02 -0.06}$ &  75$\pm9_{-12}^{+16}$ & 48$\pm7_{-7}^{+9}$ &  0 \\
\hline
\end{tabular}
\caption{Cut flow for the signal selection on signal Monte Carlo, the corresponding cosmic-ray background, the multi-jet background estimation from the ABCD method (described in Section 7) and the data; the yields are normalized to an integrated luminosity of 1.9 \textrm{fb$^{-1}$}. The first uncertainties are statistical and the second systematic.}
\label{tab:cutflowABCD}
\end{table*} 

\begin{figure}[ht!]
\centering
\includegraphics[width=70mm]{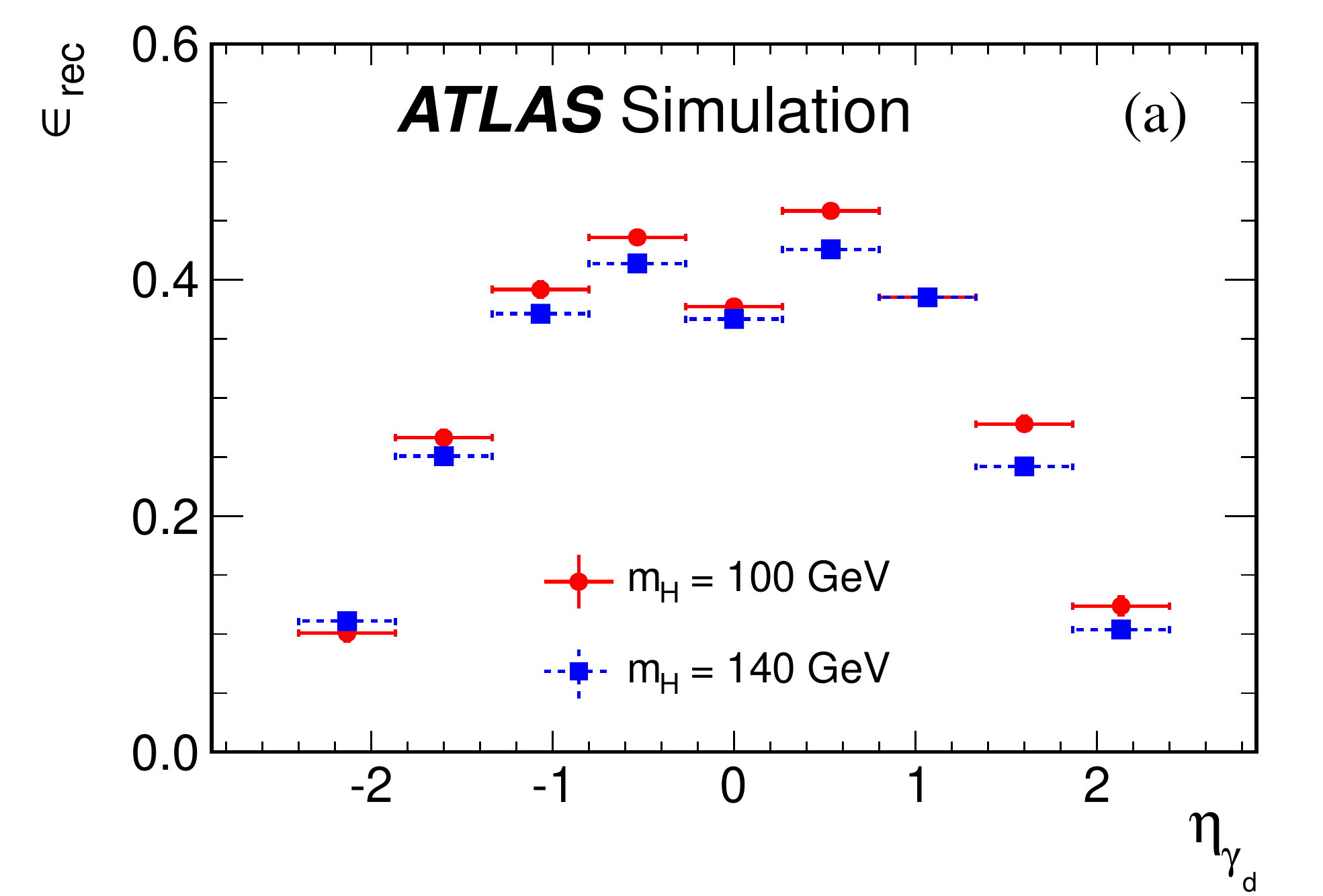}
\includegraphics[width=70mm]{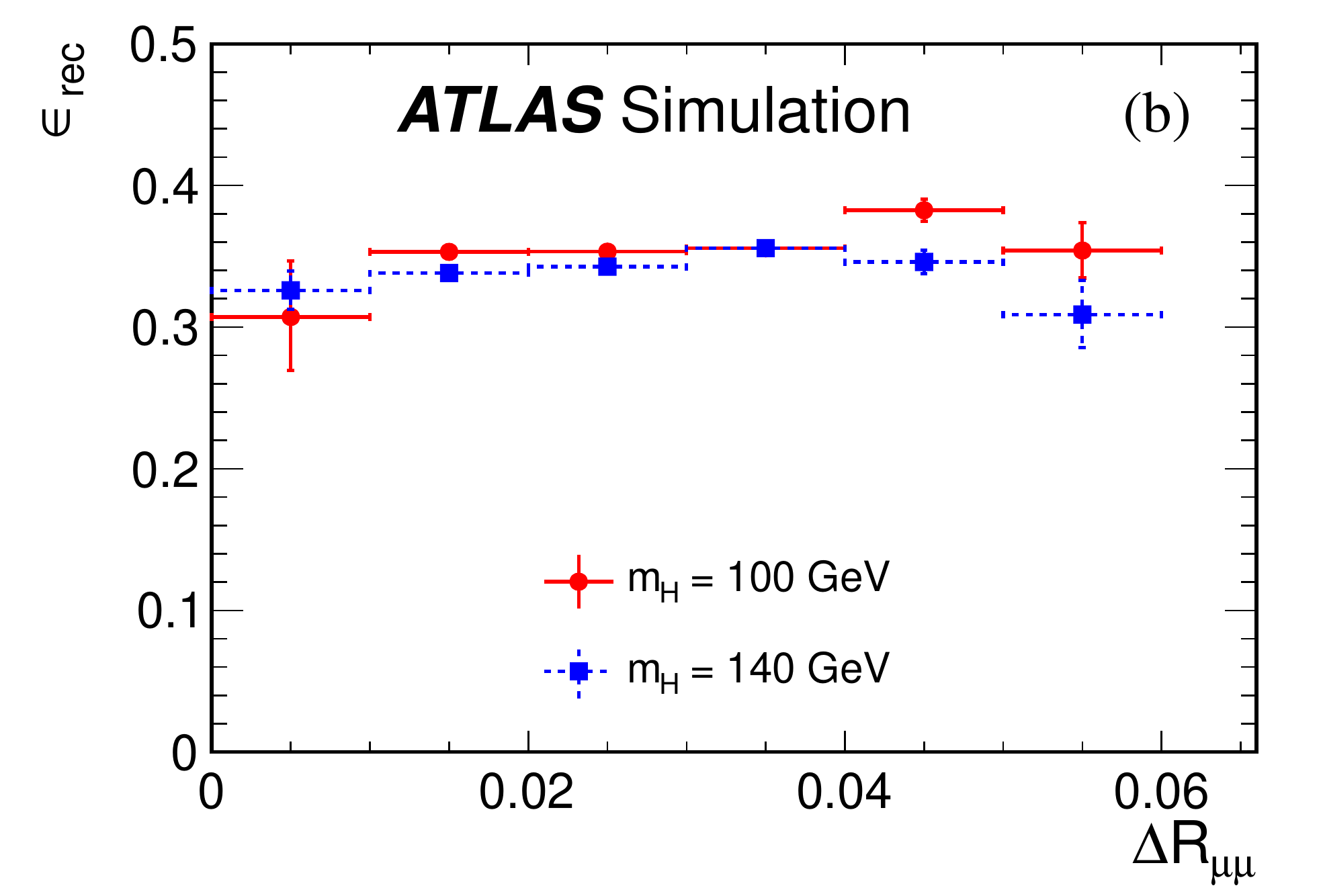}
\includegraphics[width=70mm]{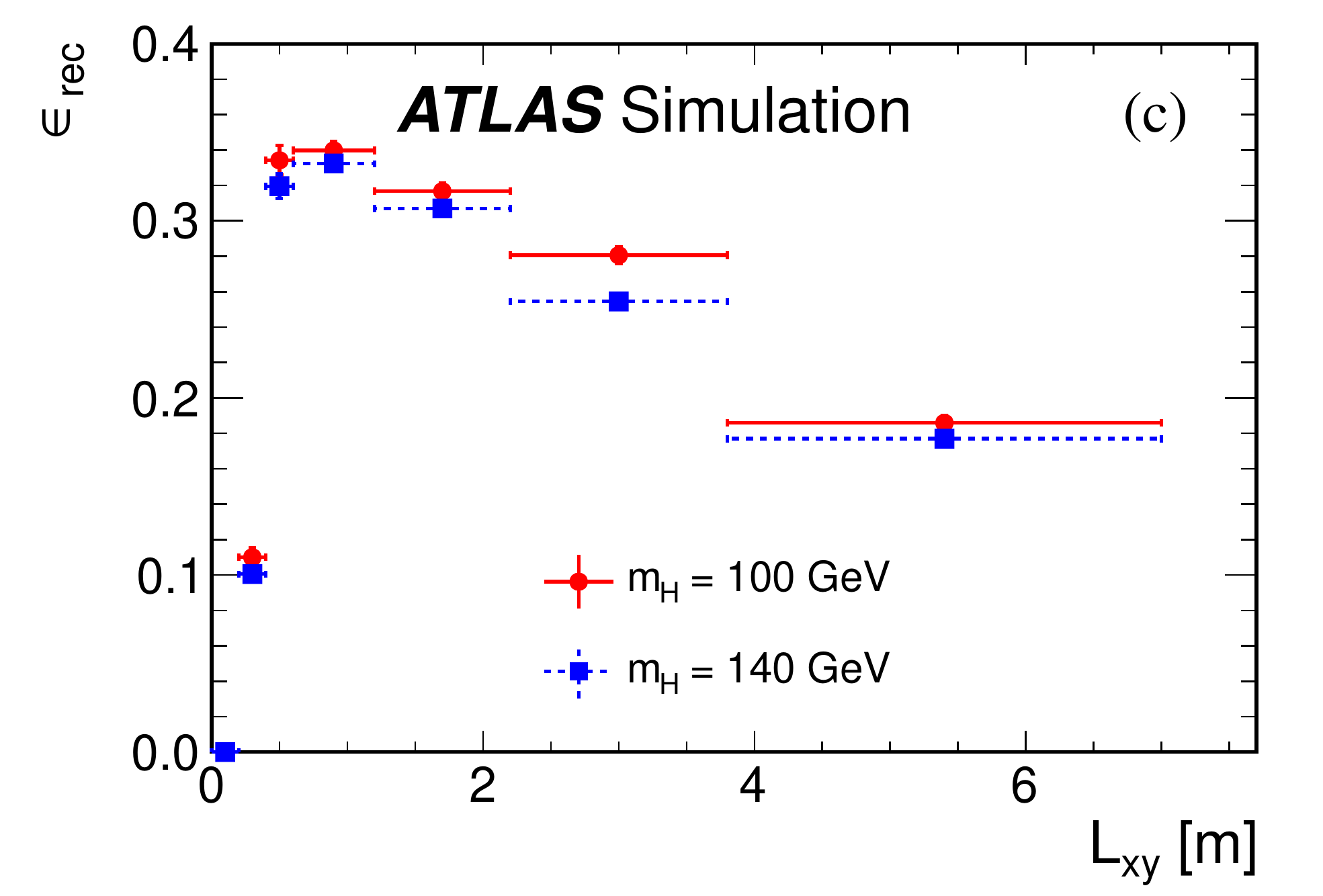}
\caption{\gammad reconstruction efficiency $\varepsilon_{\textrm\recf}$ as a function (a) of $\eta$, (b) of \mydelr and (c) of the transverse decay length of the \gammad for \higgsl and \higgsh and for the mean lifetimes given in Table~\ref{tab:param}. The reconstruction efficiency is defined as the number of \gammad passing the offline selection divided by the number of \gammad in the spectrometer acceptance (\mbox{\absetaleq~2.4}) with both muons having \pt~$\geq $ 6 \GeV. The uncertainties are statistical only.}
\label{fig:Efficiencies}
\end{figure}

The resulting single \gammad reconstruction efficiency for the mean lifetimes given in Table~\ref{tab:param} is shown in Fig.~\ref{fig:Efficiencies} as a function of $\eta$, the \mydelr separation of the two muons from the \gammad decay and the decay length in the transverse plane, \textrm{$L_{xy}$}, of the \gammad. The efficiency is defined as the number of \gammad passing the offline selection divided by the number of \gammad in the spectrometer acceptance (\mbox{\absetaleq~2.4}) with both muons having \pt~$\geq $ 6 \gev. The low reconstruction efficiency at very short \textrm{$L_{xy}$} is a consequence of the \SumpT cut. 

\begin{figure}[ht!]
\centering
\includegraphics[width=70mm]{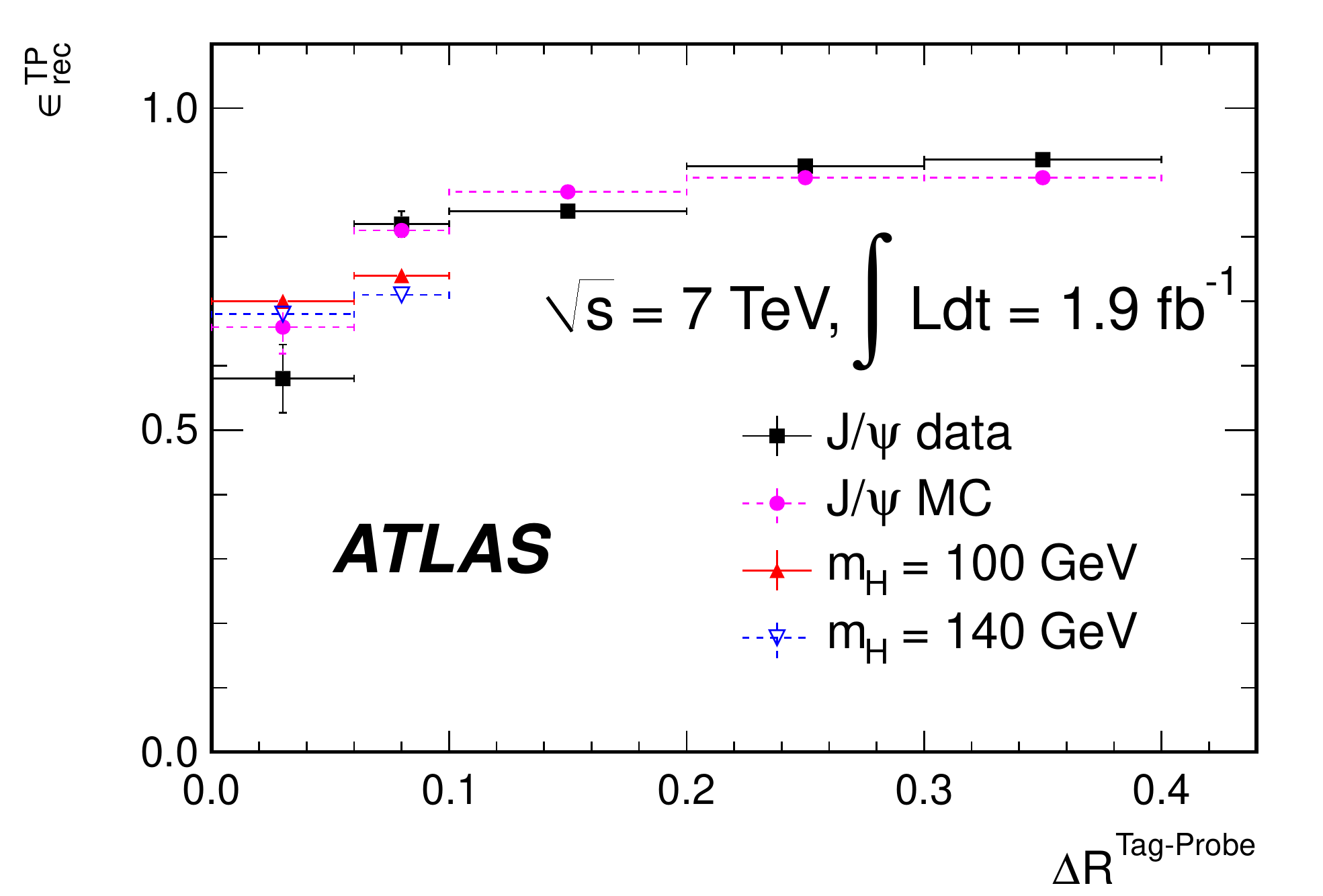}
\caption{Tag-and-probe reconstruction efficiency $\varepsilon$$_{\textrm\recss}^{\textrm\TPss}$ as a function of the \mydelr between the two muons, evaluated on a sample of \Jmumu~from collision data and a corresponding sample of Monte Carlo events. The $\varepsilon$$_{\textrm\recss}^{\textrm\TPss}$ for the signal Monte Carlo, evaluated with a similar tag-and-probe method, is also shown. The uncertainties are statistical only.}
\label{fig:effi-DR-Reco}
\end{figure}

The systematic uncertainty on the reconstruction efficiency is evaluated using a tag-and-probe method by comparing the reconstruction efficiency 
$\varepsilon$$_{\textrm\recf}^{\textrm\TPf}$ for \Jmumu~samples from collision data and \Jmumu~Monte Carlo simulation. The tag-and-probe definitions and the cut  on $\Delta R  \leq 0.1$ between the two muons  are the same as in Section 5. To measure the reconstruction efficiency the ID probe track is associated with a MS-only muon track, different from the one associated with the tag. The result is shown in Fig.~\ref{fig:effi-DR-Reco}. \\
The relative difference between the result obtained from the \Jmumu~data and the \Jmumu~ Monte Carlo sample in the same range of $\Delta R \leq 0.1$, as for the signal, is taken as the systematic uncertainty on the reconstruction efficiency and amounts to 13$\%$.

\begin{figure}[ht!]
\centering
\includegraphics[width=70mm]{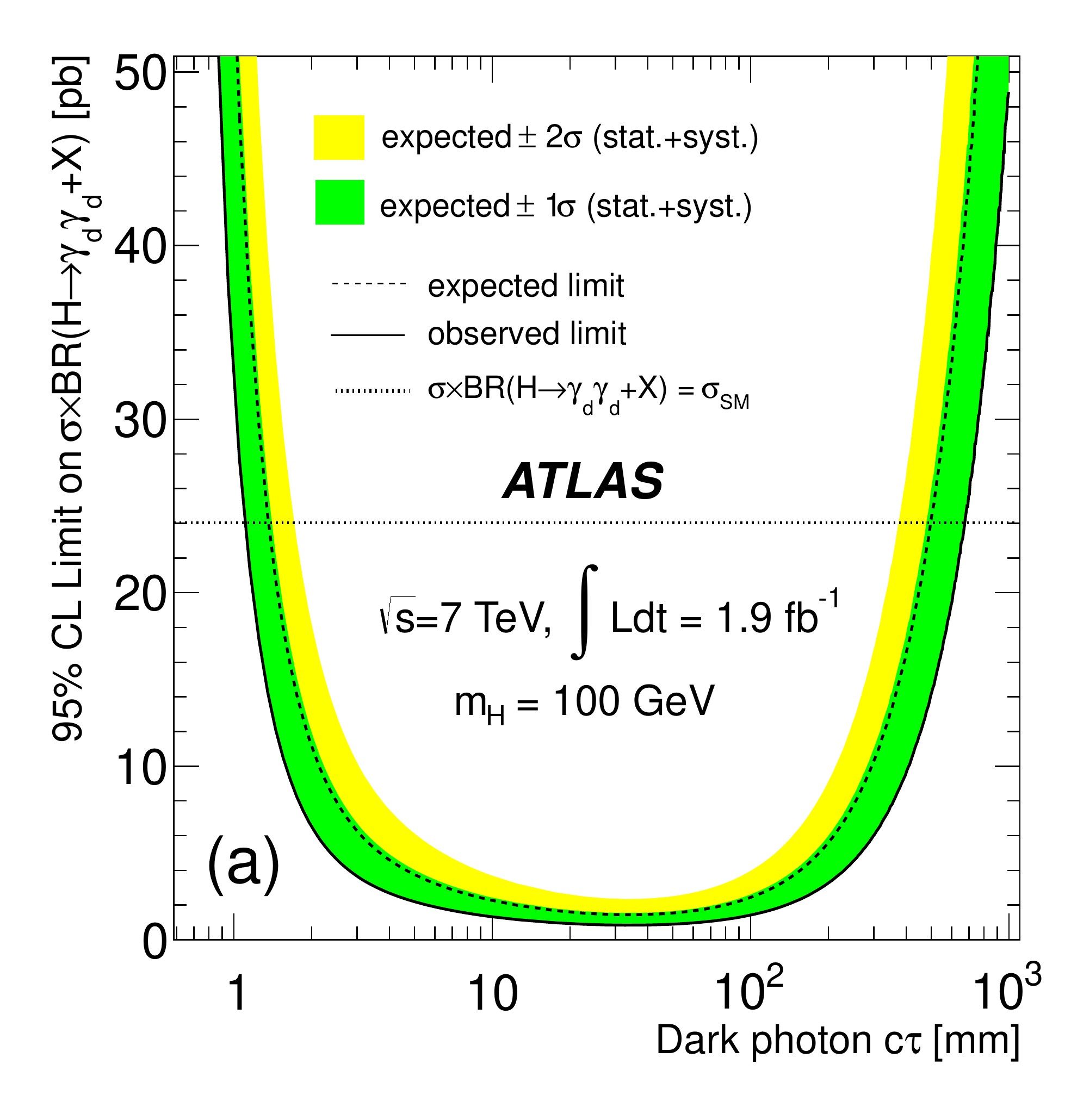}
\includegraphics[width=70mm]{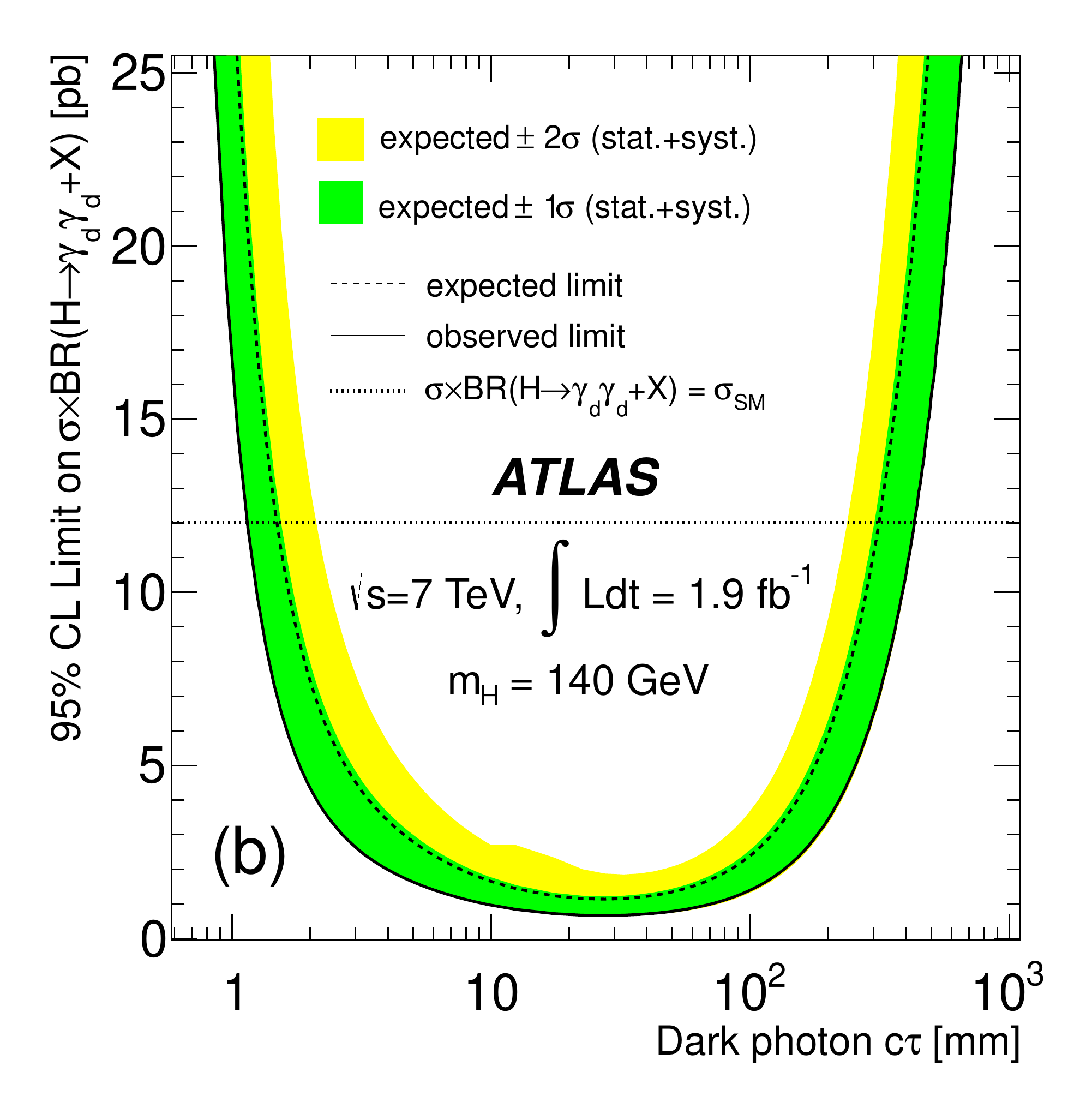}
\caption{ The 95$\%$ upper limits on the \sigmabr for the process \higgsd as a function of the dark photon \ctau for the benchmark sample with (a) \higgsl and with (b) \higgsh. The expected limit is shown as the dashed curve and the solid curve shows the observed limit. The horizontal lines correspond to the Higgs boson SM cross sections at the two mass values.}
\label{fig:CLboth}
\end{figure}

%
\vspace{0.4cm}
\maybesection{7. Multi-jet and cosmic-ray background evaluation}
\vspace{0.1cm}
To estimate the multi-jet background contamination in the signal region we use a data-driven ABCD method slightly modified to cope with the problem of the very low number of events in the control regions. The ABCD method assumes that two variables can be identified, which are relatively uncorrelated, and which can each be used to separate signal and background. It is assumed that the multi-jet background distribution can be factorized in the MJ  \isol~--~$|\Delta \phi|$ plane. The region A is defined by  \isol $\leq$ 5 \gev~ and $|\Delta \phi|<$ 2; the region B, defined by \isol $\leq$ 5 \gev~ and $|\Delta \phi|\geq$ 2, is the signal region. The regions C and D are the anti-isolated regions (\isol $>$ 5 \gev) and they are defined by $|\Delta \phi|<$ 2 and $|\Delta \phi|\geq$ 2, respectively. Neglecting the signal contamination in regions A, C and D (\isol $>$ 5 \gev~ or $|\Delta \phi|< $ 2) the number of multi-jet background events in the signal region can be evaluated as $N_B = N_D \times N_A/ N_C$. 
Due to the very low number of events in the control regions the values of $N_A$, $N_C$ and $N_D$ as a function of the cut on the final discriminant variable \SumpT are extracted  by modelling the \SumpT distributions with bifurcated Gaussian templates, with parameters fitted from the data in the corresponding regions, and by integrating the fitted function in the range 0 $<$ \SumpT $<$ 3 \gev. 
 The low statistics in the four regions at each step of the cut flow give rise to large fluctuations in the multi-jet background estimate. The extracted yields are $N_A = (7.1 \pm 1.5_{\textrm\stats})\cdot 10^{-3}$, $N_C = (1.81 \pm 1.0_{\textrm\stats})\cdot 10^{-2}$ and $N_D = (1.51 \pm 0.07_{\textrm\stats})\cdot 10^{-1}$ and the estimated number of multi-jet background events in the signal region is $N_B = 0.06 \pm 0.02_{\textrm\stats}$.
 Possible sources of systematic uncertainty related to the background estimation method are also evaluated. The functional form is changed and the procedure to estimate the number of multi-jet background events in the signal region is repeated. The difference in $N_B$ is taken as the systematic uncertainty in the modelling of the multi-jet background shape and it amounts to $^{+0.66} _{-0.06}$. The effect of possible signal leakage in the background regions is also considered and is found to be negligible. \\
The background induced by muons from cosmic-ray showers is evaluated using events collected by the trigger active when there are no collisions (empty bunch crossings). The number of triggered events is rescaled by the collision to empty bunch crossing ratio and for the active time (since the trigger in the empty bunch crossing was not active in all the runs). No events survived the requirements on the impact parameters with respect to the primary vertex ($|d_0| <$ 200 mm and $|z_0|<$ 270 mm), resulting in a cosmic-ray contamination estimate of $0^{+1.64} _{-0}$. The final yields for the different background sources are summarized in Table~\ref{tab:cutflowABCD}. 
%
\vspace{0.4cm}
\maybesection{8. Systematic uncertainties}
\vspace{0.1cm}
The following effects are considered as possible sources of systematic uncertainty:
\begin{itemize}
\item{\bf Luminosity} \\
The overall normalisation uncertainty of the integrated luminosity is $3.7\%$~\cite{LUMI1,LUMI2}. 
\item{\bf Muon momentum resolution} \\
The systematic uncertainty on the muon momentum resolution for MS-only muons has been evaluated by smearing and shifting the momenta of the muons by scale factors derived from \Zmumu~data-Monte Carlo comparison, and by observing the effect of this shift on the signal efficiency. The overall effect of the muon momentum resolution uncertainty is negligible. \item{\bf Trigger} \\
The systematic uncertainty on the single \gammad trigger efficiency, evaluated using a tag-and-probe method is $17\%$ (see Section 5).
\item{\bf Reconstruction efficiency} \\
The systematic uncertainty on the reconstruction efficiency, evaluated using a tag-and-probe method for the single \gammad reconstruction efficiency, is $13\%$ (see Section 6).
\item{\bf Effect of pile-up} \\
The systematic uncertainty on the signal efficiency related to the effect of pile-up is evaluated by comparing the number of signal events after imposing all the selection criteria on the signal Monte Carlo sample increasing the average number of interactions per crossing from $\approx 6$ to $\approx 16$. This systematic uncertainty is negligible. 
\item{\bf Effect of \SumpT cut} \\
Since the \SumpT cut could affect the minimum \ctau value that can be excluded, the effect of this cut on the signal Monte Carlo has been studied. A variation of $10\%$ on the \SumpT cut results in a relative variation of $<$1$\%$ on the signal, which can therefore be neglected.
\item{\bf Background evaluation} \\
The systematic uncertainties that can affect the background estimation are related to the data-driven method used. The functional model used to fit the \SumpT distribution is varied to evaluate the systematic uncertainty in the modelling of its shape, which also includes the effect of the \SumpT cut on the background estimation. This systematic uncertainty amounts to $^{+0.66} _{-0.06}$ events. The effect of signal leakage is also negligible. 
\end{itemize} 
%
\vspace{0.4cm}
\maybesection{9. Results and interpretation}
\vspace{0.1cm}
The efficiency of the selection criteria described above is evaluated for the simulated signal samples (see Table~\ref{tab:param}) as a function of the mean lifetime of the \gammad. Using pseudo-experiments with \ctau ranging from 0 to 700 mm the number of \gammad that decay in each region of the detector is weighted by the corresponding total efficiency for that region. In this way the number of expected signal events is predicted as a function of the \gammad mean lifetime. These numbers, together with the expected number of background events (multi-jet and cosmic rays) and  taking into account the zero data events surviving the selection criteria in \intlumi, are used as input to obtain limits at the 95$\%$ confidence level ({\it CL}). The {\it CLs} method~\cite{CLspaper} is used to set 95$\%$ {\it CL} upper limits on the cross section times branching ratio (\sigmabr) for the process \higgsd. Here the branching ratio of  \gdmumu is set to $45\%$ with the \gammad mass set to 0.4 \gev, as previously discussed. The \sigmabr is given as a function of the \gammad mean lifetime, expressed as  \ctau for \higgsl and \higgsh. These limits are shown on Fig.~\ref{fig:CLboth}. Table~\ref{tab:limits} shows the ranges in which the $\gamma_d$ \ctau is excluded at the $95\%$ {\it CL} for \higgsd branching ratios of $100\%$ and $10\%$.

\renewcommand{\arraystretch}{1.2}
\begin{table}[ht!]
\small
\centering
\begin{tabular}{|c|c c|}\hline
Higgs boson mass & excluded c$\tau$ $[mm]$ &excluded c$\tau$ $[mm]$ \\
 $[\gev]$  & BR($100\%$)& BR($10\%$) \\\hline
100& 1  $\leq$ c$\tau$ $\leq$ 670 &  5 $\leq$ c$\tau$ $\leq$ 159 \\\hline
140 & 1  $\leq$ c$\tau$ $\leq$ 430 & 7 $\leq$ c$\tau$ $\leq$ 82 \\\hline
\end{tabular}
\caption{Ranges in which \gammad \ctau is excluded at $95\%$ {\it CL} for \higgsl and \higgshnos, assuming 100$\%$ and 10$\%$ branching ratio of \higgsd.}
\label{tab:limits}
\end{table} 
 
%
\vspace{0.4cm}
\maybesection{10. Conclusions}
\vspace{0.1cm}
The ATLAS detector at the LHC was used to search for a light Higgs boson decaying into a pair of hidden fermions ($f_{d2}$), each of which decays to a \gammad and to a stable hidden fermion ($f_{d1}$), resulting in two muon jets from the \gammad decay in the final state. In a \intlumi sample of $\sqrt{s} =7$ TeV proton-proton collisions no events consistent with this Higgs boson decay mode are observed. The observed data are consistent with the Standard Model background expectations. \\
Limits are set on the \sigmabr to \higgsd as a function of the long-lived particle mean lifetime for $m_{H}=$ 100~\gev~and 140~\gev~with the chosen \gammad mass that gives a decay branching ratio of 45\% for \gdmumu. Assuming the SM production rate for a 140~\gev~Higgs boson, its branching ratio to two hidden-sector photons is found to be below 10\%, at $95\%$ {\it CL},  for hidden photon \ctau in the range 7~mm $\leq$ c$\tau$ $\leq$ 82~mm. Bounds on the \sigmabr of a 126~\gev~Higgs boson may be conservatively extracted using the corresponding 140~\gev~exclusion curve.
%
\vspace{0.4cm}
\maybesection{11. Acknowledgements}
\vspace{0.1cm}
We thank CERN for the very successful operation of the LHC, as well as the
support staff from our institutions without whom ATLAS could not be
operated efficiently.

We acknowledge the support of ANPCyT, Argentina; YerPhI, Armenia; ARC,
Australia; BMWF and FWF, Austria; ANAS, Azerbaijan; SSTC, Belarus; CNPq and FAPESP,
Brazil; NSERC, NRC and CFI, Canada; CERN; CONICYT, Chile; CAS, MOST and NSFC,
China; COLCIENCIAS, Colombia; MSMT CR, MPO CR and VSC CR, Czech Republic;
DNRF, DNSRC and Lundbeck Foundation, Denmark; EPLANET and ERC, European Union;
IN2P3-CNRS, CEA-DSM/IRFU, France; GNSF, Georgia; BMBF, DFG, HGF, MPG and AvH
Foundation, Germany; GSRT, Greece; ISF, MINERVA, GIF, DIP and Benoziyo Center,
Israel; INFN, Italy; MEXT and JSPS, Japan; CNRST, Morocco; FOM and NWO,
Netherlands; BRF and RCN, Norway; MNiSW, Poland; GRICES and FCT, Portugal; MERYS
(MECTS), Romania; MES of Russia and ROSATOM, Russian Federation; JINR; MSTD,
Serbia; MSSR, Slovakia; ARRS and MVZT, Slovenia; DST/NRF, South Africa;
MICINN, Spain; SRC and Wallenberg Foundation, Sweden; SER, SNSF and Cantons of
Bern and Geneva, Switzerland; NSC, Taiwan; TAEK, Turkey; STFC, the Royal
Society and Leverhulme Trust, United Kingdom; DOE and NSF, United States of
America.

The crucial computing support from all WLCG partners is acknowledged
gratefully, in particular from CERN and the ATLAS Tier-1 facilities at
TRIUMF (Canada), NDGF (Denmark, Norway, Sweden), CC-IN2P3 (France),
KIT/GridKA (Germany), INFN-CNAF (Italy), NL-T1 (Netherlands), PIC (Spain),
ASGC (Taiwan), RAL (UK) and BNL (USA) and in the Tier-2 facilities
worldwide.

\bibliographystyle{atlasnote}

\clearpage


\begin{flushleft}
{\Large The ATLAS Collaboration}

\bigskip

G.~Aad$^{\rm 47}$,
T.~Abajyan$^{\rm 20}$,
B.~Abbott$^{\rm 110}$,
J.~Abdallah$^{\rm 11}$,
S.~Abdel~Khalek$^{\rm 114}$,
A.A.~Abdelalim$^{\rm 48}$,
O.~Abdinov$^{\rm 10}$,
R.~Aben$^{\rm 104}$,
B.~Abi$^{\rm 111}$,
M.~Abolins$^{\rm 87}$
% ...
\bigskip

% ...
$^{10}$ Institute of Physics, Azerbaijan Academy of Sciences, Baku, Azerbaijan\\
$^{11}$ Institut de F\'{i}sica d'Altes Energies and Departament de F\'{i}sica de la Universitat Aut\`{o}noma de Barcelona and ICREA, Barcelona, Spain\\
% ...
$^{20}$ Physikalisches Institut, University of Bonn, Bonn, Germany\\
% ...
$^{47}$ Fakult\"{a}t f\"{u}r Mathematik und Physik, Albert-Ludwigs-Universit\"{a}t, Freiburg, Germany\\
$^{48}$ Section de Physique, Universit\'{e} de Gen\`{e}ve, Geneva, Switzerland\\
% ...
$^{87}$ Department of Physics and Astronomy, Michigan State University, East Lansing MI, United States of America\\
% ...
$^{104}$ Nikhef National Institute for Subatomic Physics and University of Amsterdam, Amsterdam, Netherlands\\
% ...
$^{110}$ Homer L. Dodge Department of Physics and Astronomy, University of Oklahoma, Norman OK, United States of America\\
$^{111}$ Department of Physics, Oklahoma State University, Stillwater OK, United States of America\\
% ...
$^{114}$ LAL, Universit\'{e} Paris-Sud and CNRS/IN2P3, Orsay, France\\
% ...
$^{s}$ Also at Institut f\"{u}r Experimentalphysik, Universit\"{a}t Hamburg, Hamburg, Germany\\
% ...
\end{flushleft}


\end{document}